\documentstyle[equations,epsf,12pt,world_sci]{article}   

\makeatletter
\def\eqaligntwo{\stepcounter{equation}\let\@currentlabel=\theequation
\let\\=\@eqncr
$$
\tabskip\@centering \halign to \displaywidth\bgroup
  \global\@eqcnt\m@ne\hfil
  $\@lign\displaystyle{##}$\tabskip\z@skip&\global\@eqcnt\z@
  $\@lign\displaystyle{{}##}$\hfil\qquad&\global\@eqcnt\@ne
  \hfil$\@lign\displaystyle{##}$&\global\@eqcnt\tw@
  $\@lign\displaystyle{{}##}$\hfil\tabskip\@centering&
  \llap{\@lign##}\tabskip\z@skip\crcr}

\makeatother

\def\bfp{\hbox{\boldmath $p$}}
\def\bfk{\hbox{\boldmath $k$}}
\def\bfr{\hbox{\boldmath $r$}}
\def\bfA{\hbox{\boldmath $A$}}
\def\Re{\mathop{\rm Re}\nolimits}
\def\Im{\mathop{\rm Im}\nolimits}

\def\coeff#1#2{{\textstyle{#1\over #2}}}
\def\half{\coeff{1}{2}}

\def\pmb#1{\setbox0=\hbox{$#1$}%
\kern-.025em\copy0\kern-\wd0
\kern.05em\copy0\kern-\wd0
\kern-.025em\raise.0433em\box0}

\def\beq{\begin{equation}}
\def\eeq{\end{equation}}

\def\yangmills{\vcenter{\hbox to 48pt{\hss\vbox to 20pt{\vss
\vss}\hss}}}

\def\ymdash{\vcenter{\hbox to 48pt{\hss\vbox to 20pt{\vss
\vss}\hss}}}

\def\ymobd{\vcenter{\hbox to 48pt{\hss\vbox to 20pt{\vss
\vss}\hss}}}

\def\ymtbd{\vcenter{\hbox to 48pt{\hss\vbox to 20pt{\vss
\vss}\hss}}}

\def\ymloop{\vcenter{\hbox to 0.3in{\hss\vbox to 0.4in{\vss
\vss}\hss}}}

\def\ymloopbd{\vcenter{\hbox to 0.3in{\hss\vbox to 0.4in{\vss
\vss}\hss}}}

\def\ymblip{\vcenter{\hbox to 24pt{\hss\vbox to 0.4in{\vss
\vss}\hss}}}

\def \cases#1{
           \left \{\,\vcenter {\normalbaselines  \ialign 
           {$##\hfil $&\quad {##}\hfil \crcr #1\crcr }}\right .
           }

\begin{document}

\nonfrenchspacing
\flushbottom

\title{SCREENING IN HIGH-${T}$ QCD\\[.5cm]
{~}}
\author{R.~JACKIW\thanks{This work is supported in part by funds 
provided by the U.S.
Department of Energy (D.O.E.) under cooperative agreement
\#DF-FC02-94ER40818.}\\[1ex]
{\small\em Center for Theoretical Physics, Department of Physics  
           and Laboratory for Nuclear Science } \\
{\small\em  Massachusetts Institute of Technology } \\
{\small\em                        Cambridge, MA ~02139~~U.S.A.}}
\maketitle
\bigskip
\begin{center}
To be published in: \\[1ex]
The Proceedings of \\
 {\it XVI National Meeting on Particles and Fields, \\
Caxambu, Brazil, October 1995\/} \\
\bigskip\bigskip
MIT-CTP \#2513\qquad\qquad January 1996
\end{center}
\vspace{1.7cm}
\setlength{\baselineskip}{2.6ex}

These days, as high energy particle colliders become unavailable for testing
speculative theoretical ideas, physicists are looking to other environments
that may provide extreme conditions where theory confronts physical reality.
One such circumstance may arise at high temperature $T$, which perhaps can be
attained in heavy ion collisions or in astrophysical settings.  It is natural
therefore to examine the high-temperature behavior of the standard model, and
here I shall report on recent progress in constructing the high-$T$ limit
of~QCD.

My presentation will be unified by the theme of {\bf screening}, a familiar
phenomenon in electrodynamical plasmas.  I shall explore how similar effects can be
described in QCD at a sufficiently high temperature (above the putative
confinement - deconfinement phase transition) so that we may speak of unconfirmed
quarks and gluons forming a plasma.  But first let me review briefly the screening
phenomena in plasmas of electromagnetically charged particles.  We begin with
Poisson's equation, which relates the scalar electric potential $\phi$ to a charge
density $\rho$.
$$
- \nabla^2 \phi = \rho
$$
For the charge density we take a statistical distribution of positive-charged ($+q$)
and negative-charged ($-q$) particles, each carrying the energy $\pm q \phi$,
respectively, and described by the same density $n$.  Then
$$
\rho = n (qe^{-q\phi/T}-qe^{q\phi/T}) 
$$
For large T, this becomes
$$
\raise1pt\hbox{$\rho~$}  
\lower3pt\hbox{$\widetilde{\scriptscriptstyle({\rm large}~T)}$}
\, \raise3pt\hbox{$-2nq^2\phi/T$}
$$
so that the Poisson equation reads
$$
-\nabla^2\phi + \left({2nq^2\over T}\right)\phi = O \ \ 
$$ 
Evidently, a screening mass $ \propto\left(nq^2 \over T \right)^{1 \over 2}$
has been induced for the electric potential $\phi$; the inverse is called the {\bf Debye
screening length}.  Again at high $T$ and for a relativistic plasma, one expects $n\sim
(1/{\rm volume})
\sim  T^3$,  hence the induced {\bf electric} screening mass is
$$
m \propto |q|T
$$
We shall see a similar result emerging in the non-Abelian theory as well.  Note that
Debye screening occurs for the electric (temporal) component of the gauge
potential.  There is no electrodynamical magnetic screening, because there are no
magnetic resources. 
$$
 {\pmb \nabla} \cdot {\bf B} = O
$$
In the non-Abelian theory, the corresponding equation involves the covariant divergence.

$$
{\pmb \nabla} \cdot {\bf B^a} = gf^{abc}
{\bf A}^b \cdot {\bf B}^c
$$
(Here $g$ is the gauge coupling constant.)  So the issue of magnetic sources is not so clear
in the Yang-Mills case, and one of the topics that we shall address later is whether in the
non-Abelian theory there exists magnetic screening.

The above argument -- it is essentially Debye's -- makes little use of field theoretical
formalism.  But to carry through analogous calculations in the
standard model, we shall begin with quantum field theory.  Let me explore how
finite-temperature calculations are performed in that context.  

When studying a field
theory at finite temperature, the simplest approach is the so-called imaginary-time
formalism.  We continue time to the imaginary interval $[0,1/iT]$ and consider
bosonic (fermionic) fields to be periodic (anti-periodic) on that interval. 
Perturbative calculations are performed by the usual Feynman rules as at zero
temperature, except that in the conjugate energy-momentum,
Fourier-transformed space, the energy variable $p^0$ (conjugate to the
periodic time variable) becomes discrete -- it is $2\pi n T$, ($n$
integer) for bosons.  From this one immediately sees that at high
temperature -- in the limiting case, at infinite temperature -- the
time direction disappears, because the temporal interval shrinks to
zero.  Only zero-energy processes survive, since ``non-vanishing
energy'' necessarily means high energy owing to the discreteness of
the energy variable $p^0 \sim 2\pi n T$, and therefore all modes with
$n \neq 0$ decouple at large $T$.  In this way a Euclidean
three-dimensional field theory becomes effective at high temperatures
and describes essentially static processes\@.\cite{ref1}

Let me repeat in greater detail. Finite-$T$, imaginary-time
perturbation theory makes use of conventional diagrammatic analysis in
``momentum'' space, with modified ``energy'' variables, as indicated
above. Specifically a spinless boson propagator is
$$ 
D(p)  =  \frac{i}{p_0^2 - {\bfp}^2 - m^2}{\phantom -} \qquad \qquad
p_0   =    i\pi (2n) T    \\[0.5ex]
$$
while a spin-$\frac{1}{2}$ fermion propagator reads
$$
S(p)  =   \frac{i}{\gamma^0 p_0 - {\hbox{\boldmath $\gamma \cdot p$}} - m} \qquad\qquad
p_0   =    i\pi (2n+1) T
$$
The zero-temperature integration measure $\int \frac{d^4 p}{(2
\pi)^4}$ becomes replaced by\break $iT \sum^{\infty}_{n=-\infty} \int
\frac{d^3 p}{(2 \pi)^3}$.  Thus it is seen that Bose exchange between
two $O(g)$ vertices contributes $iT \sum^{\infty}_{n=-\infty} \int
\frac{d^3 p}{(2 \pi)^3} g \frac{i}{-4 \pi^2 n^2 T^2
-{\bfp}^2-m^2} g$ where $g$ is the coupling strength. In the
large $T$ limit, all $n \neq 0$ terms (formally) vanish as $T^{-1}$ and only the
$n = 0$ term survives. One is left with $\int \frac{d^3 p}{(2 \pi)^3}
g \sqrt{T} \frac{1}{\bfp^2 +m^2} g\sqrt{T}$. This is a Bose
exchange graph in a Euclidian 3-dimensional theory, with effective
coupling $g \sqrt{T}$. Similar reasoning leads to the conclusion that
fermions decouple at large~$T$.

While all this is quick and simple, it may be physically inadequate.
First of all, frequently one is interested in non-static processes in
real time, so complicated analytic continuation from imaginary time
needs to be made before passing to the high-$T$ limit, which in
imaginary time describes only static processes.  Also one may wish to
study amplitudes where the real external energy is neither large nor
zero, even though virtual internal energies are high.

Another reason that the above may be inadequate emerges when we
consider massless fields (such as those that occur in QCD). We have
seen that the $n=0$ mode leaves a propagator that behaves as
$\frac{1}{\bfp^2}$ when mass vanishes, and a phase space of
$d^3 p$. It is well known that this kind of kinematics at low momenta
leads to infrared divergences in perturbation theory even for
off-mass-shell amplitudes --- Green's functions in massless Bosonic
field theories possess infrared divergences in naive perturbation
theory\@.\cite{ref4} Since physical QCD does not suffer from
off-mass-shell infrared divergences, perturbation theory must be
resummed.

A final shortcoming of the above limiting procedure is that it is formal: the limit is
taken before the integration/summation is carried out.  But the latter need not converge
uniformly; indeed owing to ultraviolet divergences, it may not converge at all and must be
renormalized.  As a consequence the $n \ne 0$ contributions in single Boson exchange
graphs may not decrease as $T^{-1}$. 

Thus the formal arguments for the emergence of a
3-dimensional theory at high-$T$ need be re-examined for QCD\@. Nevertheless, even if
unreliable, the arguments alert us to the possibility that 3-dimensional field theoretic
structures may emerge in the high-$T$ regime. Indeed this occurs, although not in a direct,
straightforward fashion; this will be demonstrated presently.

Here is a graphical argument to the same end discussed above: {\em
viz.}~The need to resum perturbation theory.  Consider a one-loop
amplitude $\Pi_1(p)$,
\begin{eqnarray*}
\Pi_1(p) &\equiv& \int dk ~ I_1 (p,k)   \enspace,  \\[0.5pc]
\noalign{\hbox{given by the graph in the figure.}}

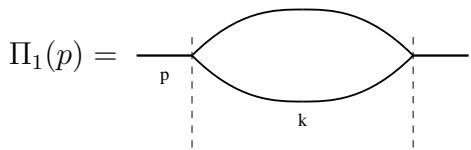
\Pi_1(p) &=&~
\vcenter{\hbox to 125pt{\vbox to 52.8pt{\vss\special{" fig1a}}\hss}}
\\[0.5pc]
&\equiv&  \qquad\quad  \int dk \,  I_1(p,k) \\[0.5pc]
\noalign{\hbox{Compare this to a two-loop amplitude $\Pi_2(p)$,}\vskip 0.5pc}
\Pi_2(p)  &\equiv&   \int dk \,  I_2 (p,k)  \enspace,  \\[0.5pc]
\noalign{\hbox{in which $\Pi_1$ is an insertion, as in the figure below.}
\vskip 0.5pc}
\Pi_2(p) &=& ~
\vcenter{\hbox to 125pt{\vbox to 52.8pt{\vss\special{" fig1b}}\hss}}
\\[0.5pc]
&\equiv& ~~~~~~~~ \, \int dk ~ I_2 (p,k)
\end{eqnarray*}
Following Pisarski,\cite{ref2}
we estimate the relative importance of $\Pi_2$ to $\Pi_1$
by the ratio of their integrands,
$$
{\Pi_2 \over \Pi_1} \sim {I_2 \over I_1} = g^2 \, {\Pi_1(k) \over k^2} ~,
$$
Here $g$ is the coupling constant, and the $k^2$ in the denominator
reflects the fact that we are considering a massless field, as in QCD\@.
Clearly the $k^2 \to 0$ limit is relevant
to the question whether the higher order graph can be neglected relative to
the lower order one.
Because one finds that for small $k$ and large
$T$, $\Pi_1(k)$ behaves as $T^2$, the ratio $\Pi_2/\Pi_1$ is $g^2 T^2 / k^2$.
As a result when $k$ is $O(gT)$ or smaller the two-loop amplitude is not
negligible compared to the one-loop amplitude.  Thus graphs with ``soft''
external momenta [$O(gT)$ or smaller] have to be
included as insertions in higher order calculations.

A terminology has arisen: graphs with generic/soft external moment [$O
(g T)\/$ where $g\/$ is small and $T\/$ is large] and large internal
momenta [the internal momenta are integration variables in an
amplitude; when $T\/$ is large they are $O (T)\/$, hence also large]
are called ``hard thermal loops\@.''\cite{ref2,ref3} Much study has
been expended on them and finally a general picture has emerged.
Before presenting general results, let us look at a specific example
--- a 2-point Green's function.

It needs to be appreciated that in the imaginary-time formalism the
correlation functions are unique and definite.  But passage to real
time, requires continuing from the integer-valued ``energy'' to a
continuous variable, and this cannot be performed uniquely.  This
reflects the fact that in real time there exists a variety of
correlation functions: time ordered products, retarded commutators,
advanced commutators, {\em etc}.  Essentially one is seeing the
consequence of the fact that a Euclidean Laplacian possesses a unique
inverse, whereas giving an inverse for the Minkowskian d'Alembertian
requires specifying temporal boundary conditions, and a variety of
answers can be gotten with a variety of boundary conditions.
Thus, when presenting results one needs to specify precisely what one is
computing.

We shall consider a correlation function for two fermionic currents,
in the 1-loop approximation.
\begin{eqnarray*}
\Pi^{\mu \nu} (x,y) 
           &=&  -i \left< j^\mu (x) j^\nu (y) \right>   \\
&=&  \int \frac{d^4 k}{(2 \pi)^4}  e^{-i k (x - y)}
           \Pi^{\mu \nu} (k)
\end{eqnarray*}
The QCD result differs from the QED result by a group theoretical
multiplicative factor, so we present high-$T\/$ results only for the
latter, in real-time, and consider the time-ordained product $\Pi^{\mu
\nu}_T\/$ as well as the retarded commutator $\Pi^{\mu \nu}_R\/$.

$\Pi^{\mu \nu}\/$ possesses a real and an imaginary part.  It is found
that at large $T\/$, the real parts of $\Pi^{\mu \nu}_T\/$ and
$\Pi^{\mu \nu}_R\/$ coincide.
$$
- {\Re} \Pi^{\mu \nu} (k) = \frac{T^2}{6} P^{\mu \nu}_2
           + \frac{T^2 k^2}{| {\bfk}^2 |}
           \left[ 1 + \frac{k^0}{2 | {\bfk} |}
           \ln \left| \frac{k^0 - | {\bfk} |}{k^0 + | {\bfk} |} \right| \right]
           \left[ \frac{1}{3} P^{\mu \nu}_1 + \frac{1}{2} P^{\mu \nu}_2 \right]
$$
where the projection operators are 
\begin{eqnarray*}
P^{\mu \nu}_1  &=&  g^{\mu \nu} - k^\mu k^\nu / k^2  \\
P^{\mu \nu}_2  &=&  
          \cases{
                 0  &  if $\mu\/$ or $\nu = 0$    \cr
                 \delta^{i j} - k^i k^j /  | {\bfk}^2 |  & otherwise \cr
           }  
\end{eqnarray*}
For the imaginary part, which is present only for space-like
arguments, different expressions are found.
\begin{eqnarray*}
- {\Im} \Pi^{\mu \nu}_R (k) 
           &\equiv&  - \pi \rho^{\mu \nu} (k)  \\
&=&  \frac{\pi k^2}{| {\bfk} |^3} \frac{k^0}{2}
           T^2 \theta (-k^2) 
           \left[ \frac{1}{3} P^{\mu \nu}_1 + \frac{1}{2} P^{\mu \nu}_2 \right]
                       \\
- {\Im} \Pi^{\mu \nu}_T (k)
           &=& \frac{\pi k^2}{| {\bfk} |^3} 
           \left[ \frac{k^0}{2} T^2 + T^3 \right]
           \theta (-k^2) 
           \left[\frac{1}{3} P^{\mu \nu}_1 + \frac{1}{2} P^{\mu \nu}_2 \right] 
\end{eqnarray*}

A unified presentation of these formulas is achieved in a dispersive
representation.  For the retarded function this reads
\begin{eqnarray*}   
\Pi^{\mu \nu}_R ({k})  
           &=&  \Pi^{\mu \nu}_{S U B} (k)
           + \int d k'_0 
           \frac{\rho^{\mu \nu} (k'_0, {\bfk})}{k'_0 - k_0 - i \epsilon}  \\
\noalign{\hbox {while the time-ordered expression is}}
\Pi^{\mu \nu}_T (k)  
           &=&  \Pi^{\mu \nu}_{S U B} (k)
           + \int d k'_0 
           \frac{\rho^{\mu \nu} (k'_0, {\bfk})}{k'_0 - k_0 - i \epsilon}
           + \frac{2 \pi i}{e^{k_0 / T} - 1}
           \rho^{\mu \nu} (k_0, {\bfk})
\end{eqnarray*}
The dispersive expressions may also be used to give the imaginary-time formula.
$$
\Pi^{\mu \nu}_{\rm imaginary \atop time} (k) 
           = \Pi^{\mu \nu}_{S U B} (k)
           + \int d k'_0 
           \frac{\rho^{\mu \nu} (k'_0, {\bfk})}{k'_0 - 2 \pi i n T}
$$
In all the above formulas, $\Pi^{\mu \nu}_{S U B} \/$ is a real
subtraction term.

Note that a universal statement about high-$T\/$ behavior can be made
only for the absorptive part $\rho^{\mu \nu} :\/$ it is $O (T^2)\/$.
This also characterizes $\Pi^{\mu \nu}_R\/$, but $\Pi^{\mu \nu}_T\/$
possesses an additional $O (T^3)\/$ imaginary part, which is seen to
arise from the additional term in $\Pi^{\mu \nu}_T\/$ involving the
bosonic distribution function $\frac{1}{e^{k_0 / T} - 1}\/$.  Finally,
the $\Pi^{\mu \nu}_{\rm imaginary \atop time}\/$ amplitude has a
temperature behavior determined by its external ``energy'' $= 2 \pi i
n T\/$.  If this is replaced by a fixed $k_0\/$ ($T\/$-independent) or
if only the $n = 0\/$ mode is considered, then one may assign an $O
(T^2)\/$ behavior to this quantity as well.

In conclusion, we assert that the 2-point correlation function behaves
as $O (T^2)\/$, where it is understood that this statement is to be
applied to the retarded amplitude, or to the imaginary time amplitude
with its ``energy'' argument continued away from $2 \pi i n
T\/$\@.\cite{ref9}

Similar analysis has been performed on the higher-point functions and
this work has culminated with the discovery (Braaten, Pisarski,
Frenkel, Taylor)\cite{ref2,ref3,ref5} of a remarkable simplicity in
their structure.  To describe this simplicity, we do not discuss the
individual $n\/$-point functions, but rather their sum multiplied by
powers of the vector potential, {\em viz.}~we consider the generating
functional for single-particle irreducible Green's functions with
gauge field external lines in the hard thermal limit.  (Effectively,
we are dealing with continued imaginary-time amplitudes.)  We call
this quantity $\Gamma_{\rm H T L} (A)\/$ and it is computed in an
$SU(N)$ gauge theory containing $N_F$ fermion species of the
fundamental representation.  $\Gamma_{\rm H T L}\/$ is found (i) to be
proportional to $(N+{1\over2} N_F)$, (ii) to behave as $T^2$ at high
temperature, and (iii) to be gauge invariant.
\def\myfig#1{
\vcenter{\hbox to 52.8pt{\vbox to 54.9pt{\vss\special{" #1}}\hss}}}
\begin{eqnarray*}
\myfig{2 0 doit}~+\myfig{3 -90 doit}+\myfig{4 45 doit}
           &+&  \myfig{5 90 doit}
           ~+ \cdots  \\
\noalign{\vskip-1ex}
&&\hskip2ex =  (N + {\textstyle{1\over2}} N_F) \, {g^2 T^2 \over 12\pi} \,
           \Gamma_{\rm HTL} (A) \\
\noalign{\vskip1ex}
\Gamma_{\rm HTL} (U^{-1} \, A \, U + U^{-1} \, dU)  \hskip-4ex
&&\hskip2ex =
           \Gamma_{\rm HTL} (A) 
\end{eqnarray*}
(Henceforth $g\/$, the coupling constant, is scaled to unity.)
A further kinematical simplification in
$\Gamma_{\rm HTL}$ has also been established.
To explain this we define two light-like four-vectors
$Q^\mu_{\pm}$
depending on a unit three-vector $\hat{q}$,
pointing in an arbitrary direction.
$$
Q^\mu_\pm  =  {1\over\sqrt{2}} (1,\,\pm \hat{q}) 
$$
\begin{eqnarray*}
\hat{q} \cdot \hat{q}  = 1  \enspace,  \qquad
           Q^\mu_\pm Q_{\pm \mu}  &=&  0 \enspace,  \qquad
           Q^\mu_\pm Q_{{\mp} \mu}  = 1
\end{eqnarray*}
Coordinates and potentials are projected onto $Q_\pm^\mu$.
$$
x^\pm \equiv x_\mu Q_\pm^\mu  \enspace,  \qquad
           \partial_\pm \equiv Q_\pm^\mu {\partial \over \partial x^\mu}   
                      \enspace,  \qquad
           A_\pm \equiv A_\mu Q_\pm^\mu
$$
The additional fact that is now known is that (iv)
after separating an ultralocal contribution from
$\Gamma_{\rm HTL}$, the remainder may be written as an average over
the angles of
$\hat{q}$
of a functional $W$ that  depends only on $A_+$;
also this functional
is non-local only on the
two-dimensional $x^\pm$ plane, and is ultralocal in the remaining directions,
perpendicular to the $x^\pm$ plane.  [``Ultralocal'' means that any potentially
non-local kernel $k(x,y)$ is in fact a $\delta$-function of the difference
$k(x,y)  \propto  \delta (x-y)$.]
$$
\Gamma_{\rm HTL} (A) = 2\pi \int d^4x ~ A^a_0 (x) A^a_0(x) +
\int d\Omega_{\hat{q}} \, W (A_+)
$$
These results are established in perturbation theory, and a perturbative
expansion of $W(A_+)$, {\it i.e.\/}~a power series in $A_+$, exhibits the
above mentioned properties.  A natural question is whether one can sum the
series to obtain an expression for $W(A_+)$.

Important progress on this problem was made when it was observed
(Taylor, Wong)\cite{ref5}
that the gauge-invariance condition
can be imposed infinitesimally, whereupon it leads
to a functional differential equation for $W(A_+)$, which is best presented as
\begin{eqnarray*}
&&{\partial \over \partial x^+} \, {\delta \over \delta A^a_+}
           \left[ W(A_+) + {\textstyle {1\over2}} 
                      \int d^4 x ~ A_+^b(x) A_+^b(x) \right]  \\
&& {\qquad\qquad\qquad}  - {\partial \over -\partial x^-} 
           \left[ A_+^a \right] + f^{abc} A_+^b
           {\delta \over \delta A_+^c}
           \left[ W(A_+) + {\textstyle {1\over2}} 
                      \int d^4 x ~ A_+^d(x) A_+^d(x) \right]
           = 0
\end{eqnarray*}
In other words we seek a quantity, call it
$$
S(A_+) \equiv W(A_+) +{1\over2} \int
d^4 x \, A_+^a (x) A_+^a(x)   \enspace, 
$$
which is a functional on a two-dimensional
manifold $\left\{ x^+, x^- \right\}$, depends on a single functional variable
$A_+$, and satisfies
$$
\partial_1 {\delta \over \delta A_1^a} S - \partial_2 A_1^a +
f^{abc} A_1^b {\delta \over \delta A_1^c} S = 0
$$
$$
{\hbox{``1''}} \equiv x^+   \enspace,  \qquad
{\hbox{``2''}} \equiv -x^-   \enspace,  \qquad
A_1^a \equiv A_+^a
$$
Another suggestive version of the above is gotten by defining $A_2^a \equiv
{\delta S \over \delta A_1^a}$.
$$
\partial_1 A_2^a - \partial_2 A_1^a + f^{abc} A_1^b A_2^c = 0
$$
To solve the functional equation and produce an expression for $W(A_+)$, we
now turn to a completely different corner of physics,
and that is Chern-Simons theory at zero temperature.

The Chern-Simons term is a peculiar
gauge theoretic topological
structure that  can be constructed in odd
dimensions, and here we consider it in  3-dimensional space-time.
$$
I_{\rm CS} \propto \int d^3 x \, \epsilon^{\alpha \beta \gamma} \,
{\rm Tr} \,
(\partial_\alpha A_\beta A_\gamma + {\textstyle {2\over3}} A_\alpha A_\beta
A_\gamma)
$$
This object was introduced into physics over a decade ago, and since that time
it has been put to various physical and mathematical uses.
Indeed
one of our originally stated motivations
for studying the Chern-Simons term
was its possible relevance
to high-temperature gauge theory\@.\cite{ref6}
Here following Efraty and Nair,\cite{ref7}
we shall employ the Chern-Simons term
for a determination of the hard
thermal loop generating functional, $\Gamma_{\rm HTL}$.

Since it is the space-time integral of a density, $I_{\rm CS}$ may be viewed
as the action for a quantum field theory
in (2+1)-dimensional space-time,
and the corresponding Lagrangian
would then be given by a two-dimensional, spatial integral
of a Lagrange density.
\begin{eqnarray*}
I_{\rm CS} &\propto&  \int dt ~ L_{\rm CS} \\
L_{\rm CS} &\propto&  \int d^2 x  \,
           \left(A_2^a \skew{4}\dot{A}_1^a + A_0^a F_{12}^a \right) 
\end{eqnarray*}
I have separated the temporal index (0) from the two spatial ones (1,2) and
have indicated time differentiation
by an over dot.
$F_{12}^a$ is the non-Abelian field strength, defined on a two-dimensional
plane.
$$
F_{12}^a = \partial_1 A_2^a - \partial_2 A_1^a + f^{abc} A_1^b A_2^c
$$
Examining the Lagrangian, we see that it has the form
$$
L \sim p \dot{q} - \lambda \, H(p,q)
$$
where $A_2^a$ plays the role of $p$, $A_1^a$ that of $q$, $F^a_{12}$
is like a Hamiltonian and $A^a_0$ acts like the
Lagrange multiplier $\lambda$, which forces the Hamiltonian
to vanish; here $A_0^a$ enforces the vanishing of $F_{12}^{\,a}$.
$$
F_{12}^{\,a} = 0
$$
The analogy instructs us how the Chern-Simons
theory should be quantized.

We postulate equal-time computation relations, like those between
$p$ and~$q$.
$$
\left[ A_1^a ({\bfr}), \, A_2^b ({\bfr}') \right]
           = i \, \delta^{ab} \delta({\bfr} - {\bfr}')
$$
In order to satisfy the condition enforced by the Lagrange multiplier, we
demand that $F_{12}^a$, operating on ``allowed'' states, annihilate them.
$$
F_{12}^a | ~~ \rangle = 0
$$

This equation can be explicitly presented in a Schr\"odinger-like
representation
for
the Chern-Simons quantum field theory, where the state is a functional of
$A_1^a$.  The action of the operators $A_1^a$ and $A_2^a$ is by multiplication
and functional differentiation, respectively.
\begin{eqnarray*}
\phantom{A_0^a} \, | ~~ \rangle &\sim&  \Psi(A_1^a) \\
A_1^a \, | ~~ \rangle &\sim&  A_1^a \, \Psi(A_1^a) \\
A_2^a \, | ~~ \rangle &\sim&  {1\over i} {\delta \over \delta A_1^a}
\, \Psi(A_1^a) 
\end{eqnarray*}
This, of course, is just the field theoretic analog of the quantum mechanical
situation where states are functions of $q$, the $q$ operator acts by
multiplication, and the $p$ operator by differentiation.
In the Schr\"odinger representation,
the condition that states be annihilated by $F_{12}^a$
$$
\left( \partial_1 A_2^{a} - \partial_2  A_1^a + f_{abc} A_1^b
A_2^c \right) \, \Big| ~~ \Big\rangle = 0
$$
leads to a functional differential equation.
$$
\left(
\partial_1 {1\over i} {\delta \over \delta A_1^a}
- \partial_2 \, A_1^a
+ f_{abc} A_1^b \, {1\over i} \, {\delta \over \delta A_1^c}
\right)
\Psi(A_1^a) = 0
$$
If we define $S$ by $\Psi = e^{iS}$ we get equivalently
$$
\partial_1 {\delta \over \delta A_1^a} S - \partial_2 A_1^a + f_{abc} A_1^b
{\delta \over \delta A_1^c} S = 0
$$
This equation comprises the entire content of Chern-Simons quantum field
theory.
$S$ is the Chern-Simons eikonal, which gives the exact wave functional owing
to the simple dynamics of the theory.
Also the above eikonal equation is
recognized to be precisely the equation
for the hard thermal loop generating functional, given above.

Let me elaborate on the connection with eikonal-WKB ideas.  Let us
recall that in particle quantum mechanics, when the wave function
$\psi (q)\/$ is written in eikonal form
$$
\psi (q) = e^{i S (q)}
$$
then the WKB approximation to $S (q)\/$ is given by the integral of
the canonical 1-form $p d q\/$
$$
S (q) = \int^q p (q') d q'
$$
where $p (q)\/$, the momentum, is taken to be function of the
coordinate $q\/$, by virtue of satisfying the equation of motion.
\begin{eqnarray*}
\frac{p^2 (q)}{2} + V (q)  &=&  E  \\
p(q)  &=&  \sqrt{2 E - 2 V (q)}
\end{eqnarray*}
Analogously, in the present field theory application, the eikonal $S
(A_1)\/$ may be written as a functional integral,
$$
S (A_1) = \int^{A_1} A^a_2 (A'_1) {\cal D} A'^a_1 
$$
where $A^a_2 (A_1)\/$ is functional of $A_1\/$ determined by the
equation of motion
$$
\partial_1 A^a_2 - \partial_2 A^a_1 + f^{a b c} A^b_1 A^b_2 = 0
$$
Since, by construction $\frac{\delta S}{\delta A^a_1} = A^a_2\/$, it
is clear that as a consequence $S\/$ satisfies the required equation.
However, we reiterate that in the Chern-Simons case there is no WKB approximation:
everything is exact owing to the simplicity of Chern-Simons dynamics.

The gained advantage for thermal physics is that ``acceptable''
Chern-Simons states, {\it i.e.\/}~solutions to the above functional
equations, were constructed long ago,\cite{ref8} and one can now take
over those results to the hard thermal loop problem.  One knows from
the Chern-Simons work that $\Psi$ and $S$ are given by a 2-dimensional
fermionic determinant, {\it i.e.\/}~by the Polyakov-Wiegman
expression.  While these are not described by very explicit formulas,
many properties are understood, and the hope is that one can use these
properties to obtain further information about high-temperature QCD
processes.  We give two applications.

The Chern-Simons information allows presenting the hard-thermal loop generating
functional as
$$
\Gamma_{\rm HTL} = {1\over2} \int d \Omega_{\hat{q}} 
[ A^a_+ A^a_- +S (A_+) + S (A_-) ] \ \ .
$$
Using the known properties of S, one can give a very explicit series expansion for
$\Gamma_{\rm HTL}\/$ in terms of powers of~$A\/$
$$
\Gamma_{\rm HTL} = {{1}\over{2!}}
           \int \Gamma^{(2)}_{\rm HTL} AA + {{1}\over{3!}} \int
           \Gamma^{(3)}_{\rm HTL} AAA + \cdots
$$
where the non-local kernels $\Gamma^{(i)}_{\rm HTL}\/$ are known
explicitly.  This power series may be used to systematize the
resummation procedure for the pertubative theory.
Here is what one does:  perturbation theory for Green's functions may
be organized with the help of a functional integral, where the
integrand contains (among other factors) $e^{i I_{\rm QCD} (A)}\/$
where $I_{\rm QCD}\/$ is the QCD action.  We now rewrite that as
$$
e^{i \left\{ I_{\rm QCD} (A) + {{m^2}\over{4 \pi}}
           \Gamma_{\rm HTL} (A) - {{m^2}\over{4 \pi}}
           \Gamma_{\rm HTL} (A) \right\} }
$$
where $m =  T \sqrt{{{N + N_F / 2}\over{3}}}\/$.  Obviously nothing
has changed, because we have merely added and subtracted the
hard-thermal-loop generating functional.  Next we introduce a loop
counting parameter $l\/$: in an $l\/$-expansion, different powers of
$l\/$ correspond to different numbers of loops, but at the end $l\/$ is
set to unity.  The resummed action is then taken to be
$$
e^{i I_{\rm resummed}} = e^{i \left\{ {{1}\over{\ell}} \left[ I_{\rm QCD} (\sqrt{\ell} A)
           + {{m^2}\over{4 \pi}} \Gamma_{\rm HTL} (\sqrt{\ell} A) \right]
           - {{m^2}\over{4 \pi}} \Gamma_{\rm HTL} (\sqrt{\ell} A) \right\} }
$$
One readily verifies that an expansion in powers of $l\/$
describes the resummed perturbation theory, and this then represents the first
application of the present Chern-Simons formalism.

For a second application, we note that even though the closed form for
$\Gamma_{\rm HTL}\/$ is not very explicit, a much more explicit formula can be
gotten for its functional derivative ${{\delta \Gamma_{\rm HTL}}\over{\delta
A^a_\mu}}\/$.  This may be identified with an induced current, which
is then used as a source in the Yang-Mills equation.  Thereby one
obtains a non-Abelian generalization of the Kubo equation, which
governs the response of the hot quark gluon plasma to external
disturbances\@.\cite{ref9}
$$
D_\mu F^{\mu\nu} =
{m^2 \over 2} j^\nu_{\rm induced}
$$
{}From the known properties   of the fermionic determinant --- hard thermal
loop
generating functional --- one can show that $j^\mu_{\rm induced}$ is given by
$$
j^\mu_{\rm induced} = \int {d \Omega_{\hat{q}} \over 4\pi} \, \left\{
Q_+^\mu \left( \vphantom{1\over1} a_-(x) - A_-(x) \right)
+ Q_-^\mu \left( \vphantom{1\over1} a_+ (x) - A_+(x) \right) \right\}
$$
where $a_{\pm}$ are solutions to the equations
\begin{eqnarray*}
\partial_+ a_- - \partial_- A_+ + [A_+, a_-] &=& 0 \\
\partial_+ A_- - \partial_- a_+ + [a_+, A_-] &=&  0
\end{eqnarray*}
Evidently $j^\mu_{\rm induced}$, as determined by the above equations, is a
non-local and non-linear functional of the vector potential $A_\mu$.

There now have appeared several alternative derivations of the Kubo
equation.  Blaizot and Iancu\cite{ref10} have analyzed the
Schwinger-Dyson equations in the hard thermal regime; they truncated
them at the 1-loop level, made further kinematical approximations that
are justified in the hard thermal limit, and they too arrived at the
Kubo equation.  Equivalently the argument may be presented succinctly
in the language of the composite effective action,\cite{ref11} which
is truncated at the 1-loop (semi-classical) level --- two-particle
irreducible graphs are omitted.  The stationarity condition on the
1-loop action is the gauge invariance constraint on $\Gamma_{\rm
HTL}\/$.  Finally, there is one more, entirely different derivation
--- which perhaps is the most interesting because it relies on
classical physics\@.\cite{ref12} We shall give the argument presently,
but first we discuss solutions for the Kubo equation.

To solve the Kubo equation, one must determine $a_{\pm}\/$ for arbitrary
$A_{\pm}\/$, thereby obtaining an expression for the induced current,
as a functional of $A_\pm\/$.  Since the functional is non-local and
non-linear, it does not appear possible to construct it explicitly
in all generality.  However, special cases can be readily handled.

In the Abelian case, everything commutes and linearizes.
One can determine $a_\pm$ in terms of $A_{\pm}$.
$$
a_\pm = {\partial_\pm \over \partial_{\mp}} \, A_{\mp}
$$
(Incidentally, this formula exemplifies the kinematical simplicity,
mentioned above, of hard thermal loops:
the nonlocality of $1/\partial_\pm$ lies entirely in the
$\left\{ x^+, x^- \right\}$
plane.)  With the above form for $a_\pm$ inserted into the
Kubo equation, the solution can be constructed explicitly.
It coincides with the results obtained by Silin long ago,
on the basis of the Boltzmann-Vlasov equation\@.\cite{ref13}
One sees that the present theory is the non-Abelian
generalization of that physics;  in particular $m$, given above,
is recognized as the Debye screening length,
which remains gauge invariant in the non-Abelian context.

It is especially interesting to emphasize that Silin did not use
quantum field theory in his derivation; rather he employed classical
transport theory.  Nevertheless, his final result coincides with what
here has been developed from a quantal framework.  This raises the
possibility that the non-Abelian Kubo equation can also be derived
classically, and indeed such a derivation has been given, as mentioned
above.

We now pause in our discussion of solutions to the non-Abelian Kubo
equation in order to describe its classical derivation.

Transport theory is formulated in terms of a single-particle
distribution function $f\/$ on phase space.  In the Abelian case,
$f\/$ depends on position $\{ x^\mu \}\/$ and momentum $\{ p^\mu \}\/$ 
of the particle.  For the non-Abelian theory it is necessary to
take into account the fact that the particle's non-Abelian charge
$\{Q^a \}\/$ also is a dynamical variable: $Q^a\/$ satisfies an evolution
equation (see below) and is an element of phase space.  Therefore, the
non-Abelian distribution function depends on $\{ x^\mu \}\/$,
$\{p^\mu \}\/$ and $\{ Q^a \}\/$, and in the collisionless
approximation obeys the transport equation ${{d}\over{d \tau}} f = 0\/$,
{\it i.e.\/}
$$
{{\partial f}\over{\partial x^\mu}} {{d x^\mu}\over{d \tau}} +
           {{\partial f}\over{\partial p^\mu}} {{d p^\mu}\over{d \tau}} +
           {{\partial f}\over{\partial Q^a}} {{d Q^a}\over{d \tau}} = 0
$$
The derivatives of the phase-space variables are given by the Wong
equations, for a particle with mass $\mu$.
\begin{eqnarray*}
{{d x^\mu}\over{d \tau}}  &=&  {{p^\mu}\over{\mu}}   \\
{{d p^\mu}\over{d \tau}}  &=&  F^{\mu \nu}_a {{d x_\nu}\over{d \tau}} Q^a \\
{{d Q^a}\over{d \tau}}  &=&  - f^{a b c}  {{d x^\mu}\over{d \tau}}
                                 A^b_\mu Q^c  \\
\end{eqnarray*}
In order to close the system we need an equation for $F^{\mu \nu}\/$.
In a microscopic description (with a single particle) one would have
$(D_\mu F^{\mu \nu})^a =
 \int d \tau Q^a {(\tau)} {{p^\nu (\tau)}\over{\mu}}
\delta^4 \big( x - x (\tau) \big)\/$
and consistency would require covariant
conservation of the current; this is ensured
provided $Q^a\/$ satisfies the equation given above.  In our
macroscopic, statistical derivation, the current is given in terms of the
distribution function, so the system of equations closes with
$$
(D_\mu F^{\mu \nu})^a  = \int d p \, d Q \, Q^a p^\nu f (x, p, Q)
$$
(One verifies that the current -- the right side of the above -- is
covariatly conserved.) The collisionless transport equation, with the
equations of motion inserted, is called the Boltzmann equation.  The closed
system formed by the latter supplemented with the Yang-Mills equation is
known as the non-Abelian Vlasov equations. To make progress, this highly
non-linear set of equations is approximated by expanding around the
equilibrium form for $f\/$,
$$
f^{\raise 1ex \hbox{$\scriptstyle \rm free$}}
           _{\hskip-0.5ex\lower 1.5ex \hbox{$
                      {\scriptstyle \rm boson \hfill}
                      \atop{\scriptstyle \rm fermion}$}}
           \propto
        \left(e^{ {{1}\over{T}} \sqrt{ \bfp^2 + \mu^2} }
                      {\scriptstyle\mp} 1\right)^{-1}
$$
This comprises the Vlasov approximation, and readily leads to the
non-Abel\-ian Kubo equation\@.\cite{ref12}

One may say that the non-Abelian theory is the minimal elaboration of
the Abelian case needed to preserve non-Abelian gauge invariance.  The
fact that classical reasoning can reproduce quantal results is
presumably related to the fact that the quantum theory makes use of
the (resummed) 1-loop approximation, which is frequently recognized
as an essentially classical effect.  Evidently, the quantum
fluctuations included in the hard thermal loops coincide with thermal
fluctuations.

Returning now to our summary of the solutions to the non-Abelian Kubo
equation that have been obtained thus far, we mention first that the
static problem may be solved completely\@.\cite{ref11} When the {\it
Ansatz\/} is made that the vector potential is time independent,
$A_\pm = A_\pm ({\bfr})\/$, one may solve for $a_\pm\/$ to find $a_\pm
= - A_\pm\/$ and the induced current is explicitly computed as
$$
{{m^2}\over{2}} j^\mu_{\rm induced} =
           {{-m^2 A^0 }\choose{ \bf 0 }}
$$
This exhibits gauge-invariant electric screening with Debye mass
$m\/$.  One may also search for localized static solutions to the Kubo
equation, but one finds only infinite energy solutions, carrying a
point-magnetic monopole singularity.  Thus there are no
plasma solitons in high-T QCD\@.\cite{ref11} Specifically, 
upon selecting the radially symmetric solution that decreases
at large distances, there arises a magnetic monopole-like singularity
at the origin.

Much less is known concerning time-dependent solutions.  Blaizot and
Iancu\cite{ref14} have made the {\it Ansatz\/} that the vector
potentials depend only on the combination $x \cdot k\/$, where $k\/$
is an arbitrary 4-vector: $A_\pm = A_\pm (x \cdot k)\/$.  Once again
$a_\pm\/$ can be determined; one finds $a_\pm = {{Q_\pm \cdot
k}\over{Q_{\mp} \cdot k}} A_{\mp}\/$, and the induced current is
computable.  For $k = \left( {{1}\atop{{\bf 0}}} \right)\/$, where
there is no space dependence (only a dependence on time is present)
one finds
$$
{{m^2}\over{2}} j^{\mu}_{\rm induced} =
           {0 \choose {- {{1}\over{3}} m^2 {\bfA} }}
$$
More complicated expressions hold with general $k\/$.  The Kubo
equation can be solved numerically; the resulting profile is a
non-Abelian generalization of a plasma plane wave.

The physics of all these solutions, as well as of other, still undiscovered
ones, remains to be elucidated, and I invite any of you to join in this
interesting task.

We see that Debye electric screening is reproduced in essentially the same form as
in an Abelian plasma (to leading order).  How about magnetic screening?  It is
important to appreciate that the above time-independent, space-independent induced
current, with ${\bf j}$ proportional to ${\bf A}$, does {\bf{not}} describe magnetic screening
because screening is determined by static configurations.   Thus we conclude that the
hard-thermal-loop limit of hot QCD does not show magnetic screening.  Indeed it appears that
if one proceeds perturbatively, beyond the resummed perturbation expansion of hard thermal
loops, no direct evidence for magnetic screening can be found. 

However, there is indirect evidence:  although the hard thermal loop resummation
cures some of the perturbative infrared divergences, as one calculates to higher
perturbative orders, they reappear essentially due to the non-linear interactions
between electric (temporal) and unscreened magnetic (spatial) degrees of freedom as well as
among the magnetic degrees of freedom due to their self-interaction.  (Such
interactions are absent in an Abelian theory.)  Consequently it is believed that
non-perturbative magnetic screening arises in the non-Abelian theory, and it is
recalled that, as mentioned in the Introduction, there is something akin to a
magnetic source in Yang-Mills theory. 

Another qualitative argument can be offered to make plausible the idea that a
magnetic mass should arise.  Although I have argued that high-temperature
dimensional reduction from four to three dimensions can not be carried out reliably
for a gauge theory, one may speculate that there is {\bf some} truth in the idea,
when restricted to magnetic (spatial) components of the non-Abelian potential.  So
one is led, as preliminary to studying the full QCD problem, to an analysis of
three-dimensional Euclidean Yang-Mills theory at zero temperature.  One quickly discovers
that infrared divergences are present in perturbation theory for this model as
well, so here again arises the question of a dynamically induced mass.  In three
dimensions, the coupling constant squared $g^2_{(3)}$ carries dimensions of mass. 
(Recall that in a high-temperature reduction 
$g_{(3)}$
is related to the four-dimensional coupling $g$ by 
$g_{(3)} = g \sqrt{T}$.)  Therefore it is plausible that three-dimensional Yang-Mills theory
generates an O$(g^2_{(3)})$ mass, which eliminates its perturbative infrared
divergences and suggests the occurrence of an 
O$(g^2 T)$ magnetic mass in the four-dimensional theory at high $T$.  Unfortunately thus far
no analysis of the three-dimensional Yang-Mills model has led to a proof of such
mass generation. 

Since the mass is not seen in perturbative expansions, even resummed
ones, one attempts a non-perturbative calculation, based on a gap equation.  Of
course an exact treatment is impossible; one must be satisfied with an approximate
gap equation, which effectively sums a large, but still incomplete set of graphs. 
At the same time, gauge invariance should be maintained; gauge non-invariant
approximations are not persuasive.

Deriving an approximate, but gauge invariant gap equation is most
efficiently carried out in a functional integral formulation.  We begin
by reviewing how a one-loop gap equation is gotten from the functional
integral, first for a non-gauge theory of a scalar field $\varphi$,
then we indicate how to extend the procedure when gauge invariance is to
be maintained for a gauge field $A_\mu$.

Consider a self-interacting scalar field theory
(in the Euclidean formulation)
whose potential $V(\varphi)$ has no quadratic term,
so in direct perturbation theory one
may encounter infrared divergences, and one enquires whether a mass is
generated, which would cure them.
\begin{eqnarray*}%
{\cal L} &=& \half \partial_\mu \varphi \partial^\mu \varphi +
V(\varphi) \\
V(\varphi) &=& \lambda_3 \varphi^3 + \lambda_4 \varphi^4 + \ldots
\label{eq:1}
\end{eqnarray*}%
The functional integral for the Euclidean theory involves
the negative exponential of the action $I = \int {\cal L}$.   Separating
the quadratic, kinetic part of $I$, and expanding the exponential of the
remainder in powers of the field yields the usual loop expansion. As mentioned
earlier, the loop expansion may be systematized by introducing a
loop-counting parameter $\ell$ and considering 
$e^{-{1\over\ell} I (\sqrt{\ell} \varphi)}$:  the power series in $\ell$
is the loop expansion.  To obtain a gap equation for a possible mass
$\mu$, we proceed by adding and subtracting $I_\mu = {\mu^2 \over 2}
\int \varphi^2$, which of course changes nothing.
$$
I = I + I_\mu - I_\mu
\label{eq:2}
$$
Next the loop expansion is reorganized by expanding $I + I_\mu$ in the
usual way, but taking $-I_\mu$ as contributing    at one    loop higher.
This is systematized as in the hard-thermal-loop application  with an effective
action,
$I_\ell$,
containing the loop counting  parameter $\ell$, which organizes    the
loop expansion    in the   indicated manner: 
$$
I_\ell = {1\over\ell} \left( I (\sqrt{\ell} \varphi)
+ I_\mu (\sqrt{\ell} \varphi) \right) - I_{\mu} (\sqrt{\ell} \varphi)
$$
An expansion in powers of $\ell$ corresponds to a resummed series;
keeping all terms and setting $\ell$ to unity returns us to the original
theory (assuming that rearranging the series does no harm);
approximations consist of keeping a finite number of terms: the
$O(\ell)$ term involves a single loop.

The gap equation is gotten by considering
the self energy $\Sigma$ of the complete propagator.  To one-loop order,
the contributing graphs are depicted in the Figure.

$$
\Sigma = {~}
\hbox to 0pt{\kern0pt\lower8pt\hbox{$\scriptstyle{\lambda_3}$}}
\yangmills
\hbox to 0pt{\kern-7pt\lower8pt\hbox{$\scriptstyle{\lambda_3}$}}
{~} +
\hbox to 0pt{\kern8pt\lower16pt\hbox{$\scriptstyle{\lambda_4}$}}
\ymloop
-
\hbox to 0pt{\kern8pt\lower16pt\hbox{$\scriptstyle{\mu^2}$}}
\ymblip
$$
\centerline{{\rm ~~Self energy resummed to one-loop order.}}
\smallskip

Regardless of the form of the exact potential, only the three- and four-
point vertices are needed at one-loop order; the ``bare'' propagator is massive
thanks to the addition of the mass term ${1\over \ell} I_\mu
(\sqrt{\ell} \varphi) = {\mu^2 \over 2} \int \varphi^2$; the last $-\mu^2$
in the Figure comes from the subtraction of the same mass term,
but at one-loop order: $-I_\mu (\sqrt{\ell} \varphi) = - \ell {\mu^2 \over
2} \int \varphi^2$.

The gap equation emerges when it is demanded 
that $\Sigma$ does not shift the mass $\mu$.
In momentum space, we require
\newpage
\begin{eqnarray*}
\Sigma(p) \biggr|_{p^2 = -\mu^2} &=& 0 \\[1.0ex]
\hbox to 0pt{\kern0pt\lower8pt\hbox{$\scriptstyle{\lambda_3}$}}
\yangmills 
\hbox to 0pt{\kern-7pt\lower8pt\hbox{$\scriptstyle{\lambda_3}$}}
{~} +
\hbox to 0pt{\kern8pt\lower16pt\hbox{$\scriptstyle{\lambda_4}$}}
\ymloop
{~} \Biggr|_{p^2 = -\mu^2} &=& \mu^2 \\
\end{eqnarray*}%
\centerline{{\rm Graphical depiction of above equation.}}
\smallskip

While these ideas can be applied to a gauge theory, it is
necessary to elaborate them so that gauge invariance is preserved.  We
shall discuss solely the three-dimensional non-Abelian Yang-Mills model (in
Euclidean formulation) as an interesting theory in its own right, and
also as a key to the behavior of spatial variables in the physical,
four-dimensional model at high temperature.

The starting action $I$ is the usual one for a gauge field.
\begin{eqnarray*}
I &=& \int d^3 x {\rm ~tr~} {\half} F^{i} F^{i} \\
F^{i} &=& {\half} \epsilon^{ijk} F_{jk} 
\label{eq:4}
\end{eqnarray*}
While one may still add and subtract a mass-generating term $I_\mu$, it is
necessary to preserve gauge invariance.  Thus we seek a gauge invariant
functional of $A_i$, $I_\mu(A)$, whose quadratic portion 
gives rise to a mass.  Evidently 
 $$
I_\mu (A) = -{\mu^2 \over 2} \int d^3 x {\rm ~tr~} A_i 
\left( \delta_{ij} - {\partial_i \partial_j \over \nabla^2} \right)
A_j + \ldots
\label{eq:7}
 $$
The transverse structure in the above equation 
guarantees invariance against
Abelian gauge transformations;
the question then remains how the quadratic term is to be completed in
order that $I_\mu(A)$ be invariant
against non-Abelian gauge transformations.
[In fact for the one-loop gap
equation only terms through $O(A^4)$ are needed.]

A very interesting proposal for $I_\mu(A)$ was given by Nair 
\cite{ref15,ref16} who also put forward the scheme for determining the
magnetic mass, which we have been describing.  By modifying in various
ways the hard thermal loop generating functional (which gives a
four-dimensional, gauge invariant but Lorentz non-invariant effective
action with a transverse quadratic term), he arrived at a gauge and
rotation invariant three-dimensional structure, which can be employed in
the derivation of a gap equation.\footnote{A gap equation for the full
gauge-field propagator, rather than just for the mass, has been put forward and
analyzed by Cornwall {\it el al.}; see {\it Phys.~Lett.}~{\bf B153}, 173 (1985)
and {\it Phys.~Rev.~D}~{\bf 34}, 585 (1986).}

Let me describe Nair's modification.  Recall that the hard-thermal-loop generating
functional, which I record here again,
$$
\Gamma_{\rm HTL} = {1\over 2} \int d \Omega_{\hat{q}} [Q^\mu_+
Q^\nu_-A^a_\mu A^a_\nu + S(Q^\mu_+A_\mu) + S (Q^\mu_-A_\mu)]
$$
is gauge invariant because $Q^\mu_\pm=(0, \pm\hat{q})$ is light-like and $S$ is the
Chern-Simons eikonal.  Before averaging over
$\Omega_{\hat{q}}$ one is dealing with a functional of only $A_\pm$; after averaging
all four components of $A_\mu$ enter.  With Nair,\cite{ref15,ref16} we observe
that another choice for $Q^\mu_\pm$ can be made, where those vectors remain
light-like, but have vanishing time component.   This is achieved when the spatial components
of $Q^\mu_\pm$ are complex and of zero length; for example:
$$
Q^\mu_+= (0,{\bf q}) \qquad \qquad Q^\mu_- = (0, {\bf q}^*)
$$
$$
{\bf q} = (-\cos\theta \cos\varphi- i\sin\varphi, -\cos\theta \sin\varphi + i \cos\varphi,
\sin\theta)=\hat{\theta} + i \hat{\varphi}
$$
Evidently ${\bf q}^2 = 0 $, $ Q^2_+ = 0$ and $Q_-=Q^{*}_+  $.  Using
these forms for
$Q^\mu_\pm$ in
$\Gamma_{\rm HTL}$ still leaves it gauge invariant.  Also it is clear that $\Gamma_{\rm HTL}$
is real and depends only on the spatial components of the vector potential. 
Hence this is an excellent candidate for $I_\mu (A) $, which therefore, following Nair, we
take it to be
$$ 
{\mu^2\over4\pi} \Gamma_{\rm HTL} (A) 
\biggr|_{{\rm evaluated}\atop{\rm as~above}}
$$

The scheme proceeds as in the scalar theory,
except that $I_\mu(A)$ gives rise not only to a mass term for the free
propagator,
but also to higher-point interaction vertices.
At one loop only the three- and four- point vertices are needed, and 
to this order the subtracting term uses only the
quadratic contribution.  Thus the gap equation reads, pictorially
$$
\left[ {~}
\ymdash ~+~ \yangmills ~+ \ymloop  + ~\ymobd~ + ~\ymtbd~ + \ymloopbd 
\right]_{p^2 = -\mu^2} = 
{~}
\hbox to 0pt{\kern8pt\lower16pt\hbox{$\scriptstyle{\mu^2}$}} \ymblip
$$

\smallskip

The first three graphs are as in ordinary Yang-Mills theory, with
conventional vertices, but massive gauge field
propagator (solid line); 
$$
D_{ij} (p) = \delta_{ij} {1\over p^2 + \mu^2}
\label{eq:5}
$$
the first graph depicting the gauge compensating ``ghost'' contribution,
has massless ghost propagators 
(dotted line)
and vertices determined by the
quantization gauge, conveniently chosen, consistent with the above form for the propagator, 
to be
$$
{\cal L}_{{\rm gauge}\atop{\rm fixing}}
= {\textstyle{1\over2}}
\pmb{\nabla} \cdot {\bf A} 
(1-\mu^2/\nabla^2)
\pmb{\nabla} \cdot {\bf A} 
\label{eq:6}
$$
The remaining three graphs arise from Nair's form for the
hard thermal loop-inspired
$I_\mu(A)$, with
solid circles denoting the new non-local vertices.  As it happens, the last graph
with the four-point vertex vanishes, while the three-point vertex reads
\begin{eqnarray*}
&& \hskip-4pt
{}^N \! V^{abc}_{ijk} ({\bf p}, {\bf q}, {\bf r}) =
\\
&& \hskip-4pt
- {i \mu^2 \, f^{abc} \over 3! ({\bf p} \times {\bf q})^2}
\, 
\left\{ {1\over3}
\left(
{{\bf p} \cdot {\bf q} \over p^2} + {{\bf r} \cdot {\bf q} \over r^2}
\right)
p_i p_j p_k
- {{\bf r} \cdot {\bf p} \over 3 r^2} 
(q_i q_j p_k + q_i p_j q_k + p_i q_j q_k )
\right\}
+ {\scriptstyle{\rm 5~permutations}}
\label{eq:11}
\end{eqnarray*}
\centerline{$p+q+r=0$}
The permutations ensure that the vertex is symmetric under the exchange
of any pair of index sets $(a~i~p), (b~j~q), (c~k~r)$.  Inverse powers of momenta signal the
non-locality of the vertex.  [We discuss the $SU(N)$ theory, with structure constants
$f^{abc}$.]

The result of the computation is
\begin{eqnarray*}
{}^N \! \Pi^{ab}_{ij} &=& \delta^{ab} \Pi_{ij}^N \\
\Pi_{ij}^N &=& \Pi_{ij}^{YM} + \overline{\Pi}_{ij}^N
\end{eqnarray*}
$\Pi_{ij}^{YM}$ is the contribution from the first three Yang-Mills
graphs and $\overline{\Pi}_{ij}^N$ sums the graphs from $I_\mu(A)$.
The reported results are \cite{ref17}
\begin{eqnarray*}
\hspace*{-.4in}
\Pi_{ij}^{YM} (p) &=& N (\delta_{ij} - \hat{p}_i \hat{p}_j)
\left[
\left( {- 13 p^2 \over 64 \pi \mu} + {5 \mu \over 16\pi} \right)
{2\mu \over p} \tan^{-1} {p \over 2\mu} - {\mu \over 16\pi} - {p \over 64}
\right] \\
&& \hbox{\qquad} + N \hat{p}_i \hat{p}_j
\left[
\left(
{p^2 \over 32 \pi \mu} + {\mu \over 8\pi} \right)
{2\mu \over p} \tan^{-1}{p \over 2\mu} + {\mu \over 8\pi} - {p \over 32}
\right] \label{eq:013} \\
\overline{\Pi}_{ij}^N (p)
&=& N \left( \delta_{ij} - \hat{p}_i \hat{p}_j \right)
\left[
\left( {3 p^2 \over 64 \pi \mu} + {3\mu \over 16\pi} \right)
{2\mu \over p} \tan^{-1} {p \over 2\mu}
- {p^2 \over 8\pi \mu} 
\left( {\mu^2 \over p^2} + 1 \right)^2 
{\mu \over p} \tan^{-1} {p \over \mu}
+ {\mu \over 16\pi} + {\mu^3 \over 8\pi p^2} + {p \over 64}
\right]
\\
&& \hbox{\qquad} - N \hat{p}_i \hat{p}_j 
\left[
\left(
{p^2 \over 32 \pi \mu} + {\mu \over 8\pi} \right)
{2\mu \over p} \tan^{-1} {p \over 2\mu} +
{\mu \over 8\pi} - {p \over 32} \right]
\label{eq:9}
\end{eqnarray*}
The Yang-Mills contribution $\Pi^{YM}_{ij}$ 
is not separately gauge-invariant (transverse) owing to the massive
gauge propagators.
[At $\mu = 0$, $\Pi^{YM}_{ij}$ reduces to the standard result
\cite{ref18}:
$N (\delta_{ij} - \hat{p}_i \hat{p}_j ) \left( - {7 \over 32} p \right)$.]
The longitudinal terms in 
$\Pi^{YM}_{ij}$ are canceled by those in 
$\overline{\Pi}^N_{ij}$, so that the total is transverse.
$$
\Pi_{ij}^N (p) = N \left( \delta_{ij} - \hat{p}_i \hat{p}_j \right)
\left[
\left(
{- 5 p^2 \over 32 \pi \mu} + {1 \over 2\pi} \mu \right)
{2 \mu \over p} \tan^{-1} {p \over 2\mu}
- {p^2 \over 8 \pi \mu} \left(  {\mu^2 \over p^2} + 1 \right)^2 
{\mu \over p} \tan^{-1} {p \over \mu} + {\mu^3 \over 8 \pi p^2} \right]
\label{eq:015}
$$
[Dimensional regularization is used to avoid divergences.]

Before proceeding, let us note the analytic structures in the above
expressions, which are presented for Euclidean momenta, but for the gap equation have to be
evaluated at the Minkowski value $p^2 = - \mu^2 < 0$.  
Analytic continuation for the inverse tangent is provided by 
$$
{1\over x} \tan^{-1} x = {1 \over 2 \sqrt{-x^2}}
\, \ln \, {1 + \sqrt{-x^2} \over 1 - \sqrt{-x^2}}
\label{eq:16}
$$
Evidently $\Pi_{ij}^N(p)$ possesses threshold singularities,
at various values of $-p^2$.  

There is a singularity at $p^2 = -4 \mu^2$ 
(from $\tan^{-1} {p\over 2\mu}$)
arising because the graphs in the Figure containing massive propagators
describe the exchange of two massive gauge ``particles''. Moreover, there
is singularity at $p^2 = -\mu^2$ (from $\tan^{-1} {p\over\mu}$)
and also, separately in $\Pi_{ij}^{YM}$ and $\overline{\Pi}^N_{ij}$,
at $p^2 = 0$ (from the $\pm {p \over 64}, \pm {p \over 32}$ terms).
These are understood in the following way.  Even though the propagators
are massive, the non-local three-point function contains 
${1\over p^2}$, 
${1\over q^2}$, 
${1\over r^2}$
contributions, which act like massless propagators.  Thus the
threshold at $p^2 = -\mu^2$ arises from the exchange of a massive line 
(propagator) together with a massless line (from the vertex).  Similarly
the threshold at $p^2 = 0$ arises from the massless lines in the vertex
(and also from massless ghost exchange).
The expressions acquire an imaginary part when the largest threshold,
$p^2=0$, is crossed:  $\Pi_{ij}^{YM}$ and $\overline{\Pi}_{ij}^N$
are complex for $p^2 < 0$.

In the complete answer, the $p^2 = 0$ thresholds cancel, and the
singularity at the $p^2 = -\mu^2$ threshold is extinguished by the
factor $({\mu^2 \over p^2} + 1)^2$.
Consequently 
$\Pi_{ij}^N$
becomes complex only for $p^2 < -\mu^2$, and is real, finite at 
$p^2 = -\mu^2$.
$$
\Pi^N_{ij} (p) \biggr|_{p^2 = -\mu^2} = 
(\delta_{ij} - \hat{p}_i \hat{p}_j) 
\, {N\mu \over 32\pi} \, 
(21 \ln 3 - 4)
\label{eq:13}
$$

From the gap equation in the last Figure, the result for the mass is \cite{ref17}
$$
\mu = {N \over 32\pi} \, (21 \ln 3 - 4) \sim  2.384 \, {N \over 4\pi}
\label{eq:14}
$$
[in units of the coupling constant $g^2_{(3)}$ (or $g^2 T$), which has been
scaled to unity].

Before accepting this plausible answer  for $\mu$, it is
desirable     to assess higher order corrections, for example two-loop
contributions.  Unfortunately, an estimate \cite{ref17} indicates that
79 graphs have to be evaluated, and the task is formidable.

An alternative test for the reliability of the above
approach and for assessing the stability of the result against
corrections has been proposed\cite{ref19}.

It is suggested that the gap equation  be derived with a gauge invariant
completion different from Nair's.  Rather than taking inspiration from hard
thermal loops (which after all have no intrinsic relevance to the
three-dimensional gauge theory\footnote{Recall that the hot thermal loop
generating functional is related to the Chern-Simons eikonal.
Since the Chern-Simons term {\it is\/} a three-dimensional structure,
this fact may provide a basis for establishing the relevance of the hard
thermal loop generating functional to three-dimensional Yang Mills
theory.  The point is under investigation by D.~Karabali and V.~P.~Nair.}), the following
formula for
$I_\mu$ is taken
$$
I_\mu(A) = {\mu^2} \int d^3 x {\rm ~tr~} F^{i} {1\over D^2} F^{i}
\label{eq:15}
$$
where $D^2$ is the gauge covariant Laplacian.  While ultimately there is
no {\it a priori\/} way to select one gauge-invariant completion 
 over another, we remark that expressions like the above 
appear in two-dimensional gauge theories (Polyakov gravity action,
Schwinger model) and are responsible for mass generation.
If two- and higher- loop effects are indeed ignorable, this alternative
gauge invariant completion, which corresponds to an alternative
resummation, should produce an answer close to the previously obtained one.

With the alternative $I_\mu$, the graphs are  as before, where the
propagator is still given by the previous expression.
However, the three- and four- point vertices in $I_\mu(A)$ are different.
One now finds for the non-local three-point vertex
$$
V_{ijk}^{abc} ({\bf p},{\bf q},{\bf r}) = {-i \mu^2 \over 3!}
f^{abc}
\left(
\delta_{ij}
{\bf q} \cdot {\bf r} + q_i p_j
\right)
{p_k \over p^2 q^2}
+ {\scriptstyle {\rm ~5~permutations}}
\label{eq:016}
$$
\centerline{$p+q+r=0$}

and the non-local four-point vertex reads
\begin{eqnarray*}
&& \hspace*{-.3in}
V^{abcd}_{ijkl} ({\bf p},{\bf q},{\bf r},{\bf s}) =
{- \mu^2 \over 4!} f^{abe} f^{cde}
\left\{
{1\over2} \delta_{jk} \, \epsilon_{imn} \, \epsilon_{\ell on} \, 
{p_m \over p^2} \, {s_0 \over s^2}
\right.
\\ 
&& - {1 \over 2 r^2}
\left.
\left(
{1\over4} \epsilon_{ijm} \epsilon_{k\ell m} - 
\epsilon_{imn} \epsilon_{k\ell n} 
{p_m \over p^2}
(p-r-s)_{j} 
+ \epsilon_{imn} \epsilon_{\ell on} {p_m \over p^2} {s_0 \over s^2}
(p-r-s)_j (p+q-s)_k
\right)
\right\} \\
&&
+ {\scriptstyle{\rm ~23~permutations}}
\end{eqnarray*}
\centerline{$p+q+r+s=0$}

These vertices do not affect the first three graphs in the Figure so that $\Pi^{YM}_{ij}$
is as before.  However, in the last three graphs the alternative non-local vertices
produce the following result, with the help of dimensional regularization, 
\begin{eqnarray*}
&& \hspace*{-.3in}
\overline{\Pi}_{ij} (p) = N (\delta_{ij} - \hat{p}_i \hat{p}_j)
\left(
\left(
{p^6 \over 128\pi \mu^5}
+ {p^4 \over 32\pi \mu^3}
+ {7p^2 \over 64\pi \mu}
+ {27\mu \over 64\pi}
- {\mu^3 \over 16\pi p^2} \right)
{2\mu\over p} \tan^{-1} {p \over 2\mu}
\right.\\
&&
\left.
-
\left(
{p^6 \over 32\pi \mu^5}
+ {p^4 \over 16 \pi \mu^3}
- {p^2 \over 16 \pi \mu}
+ {\mu \over 32\pi} \right)
\left( {\mu^2 \over p^2} + 1 \right)^2
{\mu \over p} \tan^{-1} {p \over \mu}
\right. \\
&&
\left.
- {p^2 \over 32 \pi \mu}
- {3\mu \over 16 \pi}
+ {49 \mu^3 \over 96 \pi p^2}
+ {\mu^5 \over 32 \pi p^4}
+ {p^5 \over 128 \mu^4}
+ {p^3 \over 32 \mu^2}
- {p \over 16}
\right)\\
&&
- N \hat{p}_i \hat{p}_j
\left(
\left(
{p^2 \over 32 \pi \mu}
+ {\mu \over 8\pi}
\right)
{2\mu \over p}
\tan^{-1} {p \over 2\mu}
+ {\mu \over 8\pi}
- {p \over 32}
\right)
\label{eq:18}
\end{eqnarray*}
A check on this very lengthy calculation is that summing it with
Yang-Mills contributions yields a transverse expression.
\begin{eqnarray*}
&& \hspace*{-.3in}
\Pi_{ij}(p) = N (\delta_{ij} - \hat{p}_i \hat{p}_j)
\left( 
\left(
{p^6 \over 128 \pi \mu^5} 
+ {p^4 \over 32\pi \mu^3}
- {3 p^2 \over 32 \pi \mu}
+ {47 \mu \over 64\pi}
- {\mu^3 \over 16 \pi p^2} \right)
{2\mu \over p} \tan^{-1} {p \over 2\mu}
\right.
\\
&&
\left.
- \left(
{p^6 \over 32 \pi \mu^5}
+ {p^4 \over 16\pi \mu^3}
- {p^2 \over 16 \pi \mu}
+ {\mu \over 32\pi}
\right)
\left( {\mu^2 \over p^2} + 1 \right)^2 
{\mu \over p} \tan^{-1} {p \over \mu}
\right.
\\
&&
\left.
- {p^2 \over 32 \pi \mu} 
- {\mu \over 4\pi}
+ {49 \mu^3 \over 96 \pi p^2}
+ {\mu^5 \over 32 \pi p^4}
+ {p^5 \over 128 \mu^4}
+ {p^3 \over 32\mu^2}
- {5p \over 64}
\right)
\label{eq:19}
\end{eqnarray*}
Another check on the powers of ${p \over \mu}$ is that the above reduces
to the Yang-Mills result at $\mu = 0$ \cite{ref18}.

Just as Nair's expression, the present formula
exhibits threshold singularities: at $-p^2 = 4 \mu^2$,
which are beyond our desired evaluation point
$-p^2 = \mu^2$;  there are also threshold singularities at $-p^2 = \mu^2$,
which are extinguished by the factor $({\mu^2 \over p^2} + 1)^2$; however, those
at $p^2 = 0$ do {\it not\/} cancel, in contrast to the previous case --- indeed
$\Pi_{ij}(p)$ diverges at $p^2 = 0$, and is complex for $p^2 < 0$.
[It is interesting to remark that the last graph in the last Figure,
involving the new four-point vertex, which vanishes in Nair's evaluation,
here gives a transverse result with unextinguished threshold singularities at
$-p^2 =
\mu^2$ and at $p^2 = 0$.  The protective factor of
$({\mu^2 \over p^2} + 1)^2$ arises when the remaining two graphs are added to
form $\overline{\Pi}_{ij}$, and these also contain
non-canceling $p^2 = 0$ threshold singularities, as does the Yang-Mills
contribution.]

Although $\Pi_{ij}(p) \biggr|_{p^2 = -\mu^2}$ is finite, it is complex
and the gap equation has no solution for real $\mu^2$, owing to
unprotected threshold singularities at $p^2 = 0$, which lead to a
complex $\Pi_{ij} (p)$ for $p^2 < 0$.

\begin{eqnarray*}
\mu &=& {N \over 32\pi}
\left( 29 {\textstyle{1\over4}} \ln 3 -  22 {\textstyle{1\over3}} \right)
\pm i N {13 \over 128} \\
|\mu| &\sim& 1.769 {N \over 4\pi}
\end{eqnarray*}

It may be that the hot thermal loop-inspired completion for the mass term is
uniquely privileged in avoiding complex values for $-\mu^2 \leq p^2 \leq 0$, but we see no
reason for this.\footnote{We note that Nair's hot thermal loop-inspired vertex
$^N \! V{ijk}^{abc}$ is less singular than the alternative $V_{ijk}^{abc}$, when any
of the momentum arguments vanish.  Correspondingly $\Pi_{ij}^N(p)$ is finite
at $p^2 = 0$, in contrast to $\Pi_{ij}(p)$ which diverges at ${1\over
p^2}$.  However, we do not recognize that this variety of singularities
at $p^2 = 0$ influences reality at $p^2 = -\mu^2$; indeed the individual
graphs contributing to $\Pi_{ij}^N$ are complex at that point, owing to
non-divergent threshold singularities at $p^2 = 0$ that cancel in the sum.}
Absent any argument for the disappearance of the threshold at $p^2=0$,
and reality in the region $-\mu^2 \leq p^2 < 0$, we should
expect that also the hot thermal loop-inspired calculation will exhibit
such behavior beyond the 1-loop order.\footnote{%
V.P.~Nair states that at the two loop level, there is evidence for
$\ln (1 + {p^2 \over \mu^2})$ terms, but it is not known whether they
acquire a protective factor of $({\mu^2 \over p^2}+1)$.}

Thus until the status of threshold singularities is clarified, the
self-consistent gap equation for a magnetic mass provides inconclusive
evidence for magnetic mass generation.  Moreover, if there exist gauge
invariant completions for the mass term, other than the hard thermal
loop-inspired one, that lead to real $\Pi_{ij}$ at $p^2 =
-\mu^2$, it is unlikely that they all would give the same $\mu$ at one
loop level, which is further reason why higher orders must be assessed.


\newpage

\begin{thebibliography}{99~}

\frenchspacing

\bibitem{ref1}
For a summary, see R. Jackiw and G. Amelino-Camelia, ``Field
Theoretical Background for Thermal Physics'' in {\it Banff CAP
Workshop on Thermal Field Theory\/}, F. Khanna, R.  Kobes,
G. Kunstatter, and H. Umezawa eds. (World Scientific, Singapore,~1994).

\bibitem{ref4}
R. Jackiw and S. Templeton, ``How Super-Renormalizable Interactions
Cure their Infrared Divergences,'' {\it Phys. Rev. D\/}
{\bf 23} (1981)~2291.

\bibitem{ref2}
R. Pisarski, ``How to Compute Scattering Amplitudes in Hot Gauge
Theories,'' {\it Physica\/} {\bf A158} (1989)~246.

\bibitem{ref3}
E. Braaten and R. Pisarski,
``Soft Amplitudes in Hot Gauge Theory: a General Analysis,''
{\it Nucl. Phys.\/} {\bf B337} (1990)~569;
J. Frenkel and J. C. Taylor,
``High-Temperature Limit of Thermal QCD,''
{\it Nucl. Phys.\/} {\bf B334} (1990)~199.

\bibitem{ref9}
R. Jackiw and V. P. Nair,
``High Temperature Response Function and the Non-Abelian Kubo Formula,''
{\it Phys. Rev. D\/} {\bf 48}  (1993)~4991.

\bibitem{ref5}
J. C. Taylor and S. M. Wong,
``The Effective Action of Hard Thermal Loops in QCD,''
{\it Nucl. Phys.\/} {\bf B346}  (1990)~115.

\bibitem{ref6}
R. Jackiw,
``Gauge Theories in Three Dimensions
($\approx$ at Finite Temperature)''
in {\it Gauge Theories of the Eighties\/},
R. Ratio and J. Lindfors, eds.
Lecture Notes in Physics {\bf 181}  (1983)~157
(Springer, Berlin).

\bibitem{ref7}
R. Efraty and V. P. Nair,
``Action for the Hot Gluon Plasma Based on the Chern-Simons Theory,''
{\it Phys. Rev. Lett.\/} {\bf 68}  (1992)~2891,
``Chern-Simons Theory and the Quark-Gluon Plasma,''
{\it Phys. Rev. D\/} {\bf 47}  (1993)~5601.

\bibitem{ref8}
D. Gonzales and A. Redlich, ``A Gauge Invariant Action for 2+1
Dimensional Topological Yang-Mills Theory,'' {\it Ann. Phys.\/} (NY)
{\bf 169}  (1986)~104; 
G. Dunne, R. Jackiw and C. Trugenberger,
``Chern-Simons Theory in the Schr\"odinger Representation,'' {\it
Ann. Phys.\/} (NY) {\bf 149} (1989)~197.

\bibitem{ref10}
J.-P. Blaizot and E. Iancu,
``Kinetic Equations for Long-Wavelength Excitation of Quark-Gluon Plasma,''
{\it Phys. Rev. Lett.\/} {\bf 70}  (1993)~3376,
``Soft Collective Excitations in Hot Gauge Theories,''
{\it Nucl. Phys.\/} {\bf B 417}  (1994)~608.

\bibitem{ref11}
R. Jackiw, Q. Liu, and C. Lucchesi,
``Hard Thermal Loops,
Static Response and Composite Effective Action,''
{\it Phys. Rev. D\/} {\bf 49}  (1994)~6787.

\bibitem{ref12}
P. Kelly, Q. Liu, C. Lucchesi, and C. Manuel, ``Deriving the Hard
Thermal Loops of QCD from Classical Transport Theory,''
{\it Phys. Rev. Lett.\/} {\bf 72}  (1994)~3461,
``Classical Transport Theory and Hard Thermal Loops in the Quark-Gluon
Plasma,'' {\it Phys. Rev. D\/} {\bf 50}  (1994)~4209.

\bibitem{ref13}
E. M. Lifshitz and L. P. Pitaevskii,
{\it Physical Kinetics\/} (Pergamon, Oxford UK,~1981).

\bibitem{ref14}
J.-P. Blaizot and E. Iancu,
``Non-Abelian Excitation of the Quark-Gluon Plasma,''
{\it Phys. Rev. Lett.\/} {\bf 72}  (1994)~3317,
``Non-Abelian Plane Waves in the Quark-Gluon Plasma,''
{\it Phys. Lett.\/} {\bf B326} (1994)~138.

\bibitem{ref15}

V.P.,~Nair,
``Hard Thermal loops on a Moving Plasma and a Magnetic Mass Term,''
{\it Phys.~Lett.\/} {\bf B352} (1995)~117.

\bibitem{ref16}
For a summary, see V.P.~Nair,
``Chern-Simons and WZNW theories and the quark-gluon plasma,''
Lectures at Mt.~Sorak symposium, Korea, June 1994.

\bibitem{ref17}
G. Alexanian and V.P.~Nair,
``A self-consistent inclusion of magnetic screening for the quark-gluon plasma,''
{\it Phys. Lett.\/} {\bf B352} (1995)~435.

\bibitem{ref18}
S. Deser, R. Jackiw and S. Templeton,
``Topologically Massive Gauge Theories,"
{\it Ann.~Phys.\/} (NY) {\bf 140} (1982) 372 (E) {\bf 185} (1988) 406.

\bibitem{ref19}
R. Jackiw and S.-Y. Pi, 
``Threshhold Singularities and the Magnetic Mass in Hot QCD" {\it Phys.~Lett.} {\bf B368},
131~(1996). 

\nonfrenchspacing
\end{thebibliography}
\end{document}